\documentclass[journal]{IEEEtran}
\ifCLASSINFOpdf
\else
\fi
\usepackage{graphicx}      
\usepackage[bibstyle=numeric,citestyle=numeric, uniquename=false,sortcites=true,maxbibnames=10, maxcitenames=10,doi=false, url=false, eprint=false, isbn=false, firstinits=true, backend=bibtex, natbib=true, date=year]{biblatex}
\usepackage{amsmath, amsthm}
\usepackage{amssymb}
\usepackage{color}
\usepackage{float}
\usepackage{algorithm}
\usepackage{booktabs}
\usepackage{enumitem}

\usepackage{array}
\usepackage{adjustbox}

\newtheorem{theorem}{Theorem}

\newtheorem{assumption}{Assumption}
\newtheorem{remark}{Remark}

\newcommand{\outcomment}[1]{\iffalse #1 \fi }
\usepackage{epstopdf}

\newcommand{\nodState}{\chi}

\definecolor{darkblue}{rgb}{0,0,0.9}
\definecolor{orange}{rgb}{1,0.5,0}
\definecolor{htmlgreen}{rgb}{0.0, 0.5, 0.0}

\usepackage{xspace}
\newcommand{\aladin}{\textsc{aladin}\xspace}
\newcommand{\admm}{\textsc{admm}\xspace}
\newcommand{\nlp}{\textsc{nlp}\xspace}
\newcommand{\nlps}{\textsc{nlp}s\xspace}
\newcommand{\opf}{\textsc{opf}\xspace}
\newcommand{\ac}{\textsc{ac}\xspace}
\newcommand{\kkt}{\textsc{kkt}\xspace}
\newcommand{\ocd}{\textsc{ocd}\xspace}
\newcommand{\sdp}{\textsc{sdp}\xspace}
\newcommand{\ieee}{\textsc{ieee}\xspace}
\newcommand{\sqp}{\textsc{sqp}\xspace}
\newcommand{\qp}{\textsc{qp}\xspace}
\newcommand{\bfgs}{\textsc{bfgs}\xspace}

\newcommand{\comm}[1]{\textcolor{black}{#1}}

\newcommand{\tNLP}{\hat{T}_{\text{\nlp}}}
\newcommand{\tQP}{\hat{T}_{\text{\qp}}}
\newcommand{\tWC}{\hat{T}_{\textsc{wc}}}
\newcommand{\nFW}{\hat{N}_{\textsc{fw}}}
\newcommand{\nBW}{\hat{N}_{\textsc{bw}}}
\newcommand{\nFWeff}{N_{\textsc{fw}}}

\bibliography{paper}

\usepackage{enumitem}

\begin{document}
%
\title{Towards Distributed OPF using ALADIN}
%
%
%

\author{Alexander Engelmann, Yuning Jiang~\IEEEmembership{Student Member,~IEEE}, Tillmann M\"uhlpfordt, Boris Houska and \\  Timm Faulwasser~\IEEEmembership{Member,~IEEE} 
	\thanks{AE, TM and TF are with the Institute for Automation and Applied Informatics,
		Karlsruhe Institute of Technology, Germany
		{\tt\small \{alexander.engelmann, tillmann.muehlpfordt, timm.faulwasser\}@kit.edu}.
	}%
	\thanks{YJ and BH are with the School of Information Science and Technology, ShanghaiTech University, Shanghai, China
		{\tt\small \{borish, jiangyn\}@shanghaitech.edu.cn}. 
	}%
	\thanks{
		BH and TF acknowledge support of this joint research by the Deutsche Forschungsgemeinschaft Grant WO 2056/4-1.
		TF acknowledges funding from the Baden-W\"urttemberg Stiftung under the Elite Program for Postdocs.
		This work was also supported by the Helmholtz Association under the Joint Initiative ``Energy System 2050 - A Contribution of the Research Field Energy''.
	}%
\thanks{Manuscript received TBD.}
}

%
%

\markboth{Submitted}
{Shell \MakeLowercase{\textit{et al.}}: Bare Demo of IEEEtran.cls for IEEE Journals}
%



\maketitle

\begin{abstract}
The present paper discusses the application of the recently proposed Augmented Lagrangian Alternating Direction Inexact Newton (\aladin) method to non-convex \ac Optimal Power Flow Problems (\opf) in a distributed fashion. In contrast to the often used Alternating Direction of Multipliers Method (\admm), \aladin guarantees locally quadratic convergence for \ac-\opf. Numerical results for 5--300 bus test cases indicate that \aladin is able to  outperform \admm and to reduce the number of iterations by about one order of magnitude. We compare \aladin to numerical results for \admm documented in the literature.
The improved convergence speed comes at the cost of increasing the communication effort per iteration. Therefore, we propose a variant of \aladin that uses inexact Hessians to reduce communication. \comm{Additionally, we provide a detailed comparison of these \aladin variants to \admm from an algorithmic and communication perspective. Moreover, we prove that \aladin converges locally at quadratic rate even for the relevant case of suboptimally solved local \nlps.}
\end{abstract}


%
\IEEEpeerreviewmaketitle

\section{Introduction}
\IEEEPARstart{O}{ptimal} power flow (\opf) problems (or variants thereof) are employed in many power system contexts to ensure stable and economic system operation. 
In presence of line congestions for example, \opf problems are used to determine/re-dispatch generator set points.
In fact, in  the German power grid, the number of these re-dispatch events increased drastically in recent years owing to the increasing penetration of renewables, phase-out of nuclear plants, and liberalized  energy markets \cite{BundesnetzagenturfuerElektrizitaet2017}.
This trend illustrates the importance of efficient and reliable \opf computations in daily grid operation.
Whereas in the past, the distribution grid level was often not considered in \opf computations, nowadays this may lead to problems as renewable generation might cause violation of voltage limits and line limits at the distribution grid level.
Including the distribution grid to \opf problems under \ac conditions can help to resolve this, yet doing so increases the problem size. 
All of the above observations have triggered significant research activity on hierarchical, respectively, distributed algorithms for \opf; i.e. algorithms that split the overall problem into a number of smaller subproblems whose parallel solution may or may not be coordinated by a central entity  \cite{Molzahn2017}.\footnote{We remark that the notions of \emph{distributed algorithms} are not unified in the context of numerical optimization for \opf problems: While in the optimization literature distributed algorithms entail a central coordinating entity \cite{Bertsekas1989}, in the context of \opf such schemes are referred to as being \emph{hierarchical} \cite{Molzahn2017}.}

Given the relevance of solving \opf problems, it is not surprising that there exists a multitude of results on distributed algorithms for \opf under \ac conditions; we refer to \cite{Molzahn2017,Capitanescu2016} for recent overviews.
One can distinguish three main lines of research: i) (ad-hoc) application of algorithms tailored to convex Nonlinear Programs (\nlps) thus in general losing convergence properties \cite{Erseghe2014a,Erseghe2015}; ii) convex relaxation of \opf  by either inner or outer approximation of the feasible set \cite{Low2014,DallAnese2013a}; and iii) application of distributed algorithms tailored to non-convex \nlps \cite{Engelmann2017,Hours2017}.
The present paper follows along iii).
Before we present our approach, we concisely review existing results for items i)-iii).

With respect to i), the set of convex algorithms directly applied to \opf ranges from the Auxiliary Problem Principle \cite{Kim1997,Hur2002}, the Predictor Corrector Proximal Multiplier Method  \cite{Kim2000}, to the popular Alternating Direction of Multipliers Method (\admm) \cite{Kim2000,Erseghe2014a}. 
A number of recent works discusses \admm in more detail, each with different foci: exhaustive simulation-based convergence analysis \cite{Erseghe2014a}, parameter update rules \cite{Erseghe2015}, applicability to large-scale grids \cite{Guo2017}.
All of these methods share the advantage that, usually, they exchange only primal variables between the subproblems, which correspond to linear consensus constraints. However, the convergence rate is at most linear \cite{Boyd2011,Bertsekas1989}.
Recently, the authors of \cite{Hong2016} presented \admm convergence results for problems with non-convex objective function of a special form (consensus and sharing problems).
However, it remains unclear whether \ac-\opf fits that form and, to the best of the authors' knowledge, there are no general convergence guarantees for \ac-\opf using \admm.

Another subbranch of i) proposes Optimality Condition Decomposition (\ocd) to solve \ac-\opf problems, see \cite{Conejo2002, Conejo2006} and \cite{Nogales2003, Arnold2007,Hug-Glanzmann2009}. This method aims to solve the first-order necessary conditions including nonlinear coupling constraints without any problem modification in a distributed fashion. In \ocd, all subproblems receive primal and dual variables from neighboring regions and consider them as fixed parameters in each local optimization. In \cite{Conejo2002}, a necessary condition for convergence to a first-order stationary point is discussed. However, to the best of our knowledge, it remains unclear whether this condition holds for arbitrary \opf problems \cite{Erseghe2014a}. Moreover, in \cite{Conejo2002}, the convergence rate of this method is shown to be linear. 

With respect to ii), convex outer approximations of the feasible set via Semi-Definite Programming (\sdp) are considered in \cite{Bai2008, DallAnese2013a,Molzahn2013a,Peng2017}; the \opf problem is mapped to a higher dimensional space wherein it becomes convex whenever a specific rank constraint is dropped. This relaxed and inflated problem can be solved using the above mentioned convex algorithms obtaining convergence guarantees. The crux of \sdp relaxations of \opf problems is that the exactness of solutions (in terms of the original non-relaxed \opf problem) can so far only be guaranteed via structural assumptions: either on technical equipment like small transformer resistances or on the grid topology, e.g. radial grids \cite{Lavaei2012,Low2014,Low2014a,Christakou2017}. 

Finally, research line iii) considers algorithms with certain convergence guarantees for non-convex problems. This includes approaches based on trust region and alternating projection methods  with convergence guarantees at linear rate \cite{Hours2017}.
A distributed approach based on interior point algorithms is proposed in \cite{Lu2018}, where the authors (similar to works on optimal control \cite{Necoara2009,TranDinh2013}) decompose certain steps in of a centralized optimization method. Hence,  \cite{Lu2018} \mbox{obtains---due} to equivalence to the corresponding centralized \mbox{method---promising} numerical results even for very large grids. 

The present paper aims at investigating the potential of the recently proposed Augmented Lagrangian Alternating Direction Inexact Newton (\aladin) method \cite{Houska2016} for \opf problems. 
Similar to \admm, \aladin solves a sequence of local optimization problems combined with a coordination step.
All computationally expensive operations (i.e. non-convex minimizations as well as function and derivative evaluations) are performed locally.
The coordination step entails solving an equality-constrained Quadratic Program (\qp) in each iteration which is computationally cheap as an equality-constrained \qp results in a linear system of equations. Motivated by the locally quadratic convergence properties of \aladin \cite{Houska2016}, we proposed its application to \opf in a preceding conference paper \cite{Engelmann2017} presenting results merely for a 5-bus problem without line limits. 

\comm{The contributions of the present paper are threefold:
	(a) We provide a  detailed investigation of the prospect of \aladin for \opf problems. To this end, we present numerical results of \aladin  for a set of widely used (\ieee) test systems ranging from 5 to 300 buses. 
	We explicitly compare our findings to \admm results presented in \cite{Erseghe2015}. 
	(b) We show how inexact Hessians  can be used to reduce the communication effort of \aladin, and provide a detailed analysis for the test systems.
	(c) Finally, we prove  quadratic convergence for the practically relevant case of suboptimal solution of the local \nlps extending the convergence analysis of \cite{Houska2016}.
}

The remainder of the paper is structured as follows: Section \ref{sec:ProblemForm} states the \ac-\opf problem. Section \ref{sec:ALADIN}, recalls \ac-\opf problem in affinely coupled separable form and revisits the \aladin algorithm. Extensive numerical case studies for \aladin and \admm are discussed Section \ref{sec:results}. Finally, Section \ref{sec::ALvsADM} compares \aladin and \admm in terms of their convergence properties and in terms of their communication effort.

\emph{Notation}
Subscripts $(\cdot)_{k,l}$ describe nodal variables, subscripts $(\cdot)_{i,j}$ denote local variables, and superscripts $(\cdot)^k$ indicate \aladin iterates.

\section{Problem Statement}
\label{sec:ProblemForm}
\subsection{Optimal Power Flow Problem}

\label{sec:ACOPF}
Consider an electrical grid at steady state described by the triple $(\mathcal{N}^0,\mathcal{G},Y)$, where $\mathcal{N}^0=\{1, \hdots, N^0\}$ is the bus set, $\mathcal{G} \subseteq \mathcal{N}^0$ is the generator set and  $Y=G+jB \in \mathbb{C}^{N^0\times N^0}$ is the bus admittance matrix. 
Neglecting shunts for simplicity,
the entries $Y_{kl} = G_{kl} + j B_{kl}$ of the bus admittance matrix are given by
\[
Y_{kl}=
\left\{
\begin{aligned}
 \displaystyle \sum_{  m \in \mathcal{N}^0 \setminus \{k\}} &y_{km},  && \mbox{if} \quad k=l, \\
-&y_{kl},  && \mbox{if} \quad k \neq l, 
\end{aligned}\right.
\]
where $y_{kl} \in \mathbb{C}$ is the admittance of the transmission line connecting buses $k$ and $l$.
One bus $r\in \mathcal{N}^0$ is specified as reference bus for the voltage angles. The \ac-\opf problem can be written as the following \nlp
\begin{subequations}
	\label{eq::ACOPF}
	\begin{align}
		&\min_{\theta,v,p,q} ~ \sum_{k \in \mathcal{G}} c_{1,k} p_{k}^2 + c_{2,k} p_{k} + c_{3,k}, \\
		&\quad \text{subject to}\;\nonumber\\ 
		\begin{split}
		 &\;v_k\sum_{l \in \mathcal{N}^0} v_l(G_{kl}\cos(\theta_{kl})+B_{kl}\sin(\theta_{kl})) = p_k - p_{k}^d  ,\label{eq::pf1}\\[0.1cm]
		&\;v_k\sum_{l \in \mathcal{N}^0} v_l(G_{kl}\sin(\theta_{kl})-B_{kl}\cos(\theta_{kl})) = q_k - q_{k}^d ,
		\end{split}	 \\[0.1cm]
		\begin{split}
		& \; \underline{p}_k \leq p_k \leq \overline{p}_k, \quad \forall k \in \mathcal{G},\label{eq::box1} \\[0.1cm]
		&\; \underline{q}_k \leq q_k \leq \overline{q}_k, \quad \forall k \in \mathcal{G}, \\[0.1cm]
		&\; \underline{v}_k \leq v_k \leq \overline{v}_k, \quad \forall k \in \mathcal{N}^0,
		\end{split}\\[0.1cm]
		&\; v_r=1\;, \;\; \theta_r=0\; \label{eq::slack}, 
	\end{align}
\end{subequations}
with $c_{1,k}>0$ and $\theta_{kl}=\theta_k-\theta_l$. In Problem~\eqref{eq::ACOPF} $v_k$ denotes the voltage magnitude, $\theta_k$ denotes the voltage angle, $p_k$ and $ q_k$ denote the active and reactive power injections,  $p_k^d$ and $q_k^d$ denote the active and reactive power demands at bus $k$. Problem~\eqref{eq::ACOPF} aims to minimize the total generation cost subject to the power flow equations~\eqref{eq::pf1}, generation and voltage bounds~\eqref{eq::box1}, and the reference constraint \eqref{eq::slack}. 

\subsection{Separable Reformulation}

\label{sec:ACOPF_separable}
We recall the reformulation of the \ac-\opf Problem~\eqref{eq::ACOPF} in affinely coupled separable form amenable to distributed optimization \cite{Engelmann2017}.

\begin{figure}
	\centering
	\includegraphics[width=0.8\linewidth]{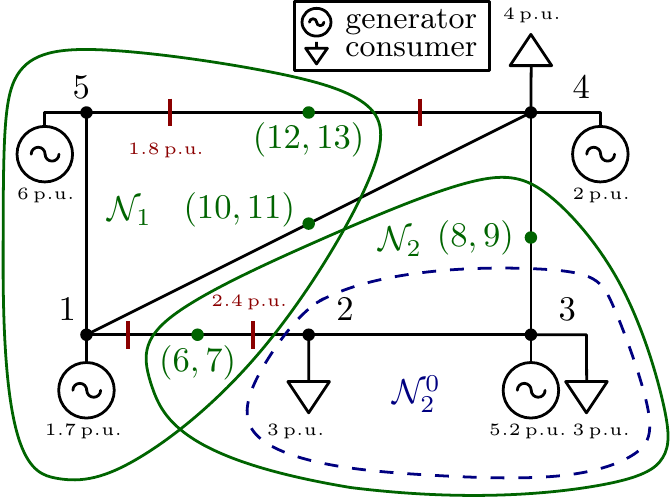}
	\caption{Decomposed 5-bus test case \cite{Li2010} with three local bus sets $\mathcal{N}_1=\{1,5,6,10,12\}$, $\mathcal{N}_2=\{2,3,7,8\}$, $\mathcal{N}_3=\{4,9,11,13\}$ (black), auxiliary bus pairs $\mathcal{A}=\{(6,7),(8,9),(10,11),(12,13)\}$ (green) and line limits depicted in red.}
	\label{fig:grid}
\end{figure}

\begin{figure} 
	\centering
	\includegraphics[width=0.85\linewidth]{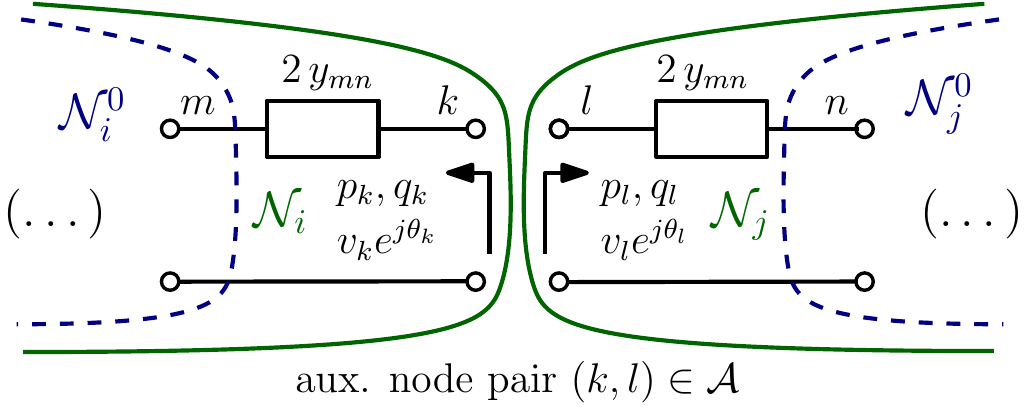}
	\caption{\comm{Coupling of two neighbored regions.}}
	\label{fig:splitting2}
\end{figure}

We begin by partitioning the bus set $\mathcal{N}^0$ into $\mathcal{R}=\{1,\dots,R\}$ (usually geographically motivated) distinct local bus sets $\mathcal{N}^0_i=\{n^{0,1}_i,\dots,n_i^{0,N^0_i}\}$.
For each bus pair $(m,n)$ located at a boundary between two local bus sets (which means $m \in \mathcal{N}^0_i$ and $n \notin \mathcal{N}^0_i$), we introduce an auxiliary bus pair $(k,l)$  in the middle of the corresponding transmission line. 
Hence, the corresponding admittances coupling bus $m$ and $k$ ($n$ and $l$ respectively)  are twice as big as the original admittance, i.e. $y_{mk} = 2\,y_{mn}$, $y_{nl} = 2\,y_{mn}$.
We couple the auxiliary buses only with buses in the interior of each region (i.e. not with each other). Thus we obtain decoupled local admittance matrices $Y_i \in \mathbb{C}^{N_i\times N_i}$ that contain all original buses and the newly introduced auxiliary buses.
Furthermore, we define enlarged local bus sets $\mathcal{N}_i=\{n^{1}_i,\dots,n_i^{N_i}\}$  containing the original local bus sets $\mathcal{N}_i^0$ and their corresponding auxiliary bus.
Fig. \ref{fig:grid} and Fig. \ref{fig:splitting2} show the decomposition procedure and the corresponding sets exemplarily for a 5-bus system.
All auxiliary bus pairs are collected in the set $\mathcal{A}$ and the enlarged local bus sets define the enlarged bus set $\mathcal{N}=\bigcup_{i \in \mathcal{R}} \mathcal{N}_i$.

%

\comm{Every bus $k \in \mathcal{N}$ is represented by
\begin{equation*}\label{eq:NodalState}
	\nodState_{k} =	\left [\, \theta_k \quad v_k\quad p_{k}\quad q_{k}\; \right]^\top \in \mathbb{R}^4.
\end{equation*}
For each region $i \in \mathcal{R}$, we stack its bus variables $\nodState_{k}$ in local vectors ${x_i= [\nodState_{n_1^i}^\top\;\dots\;\nodState_{n_{N_i}^i}^\top ]^\top \in \mathbb{R}^{n_i}}$ where $n_i=4N_i$.
The local objective functions $f_i:\mathbb{R}^{n_i}\rightarrow\mathbb{R}$ are
\begin{equation} 
\notag
f_i(x_i) := \sum_{k \in \mathcal{G}_i} c_{1,k} p_{k}^2 + c_{2,k} p_{k} + c_{3,k},
\end{equation}
where $\mathcal{G}_i=\mathcal{N}_i\,\cap\, \mathcal{G}$ denote local generator sets.
The power flow equations~\eqref{eq::pf1} and slack constraints \eqref{eq::slack} are formulated as local nonlinear equality constraints $h_i:\mathbb{R}^{n_i}\rightarrow \mathbb{R}^{n_{hi}}$. 
}

Summarizing the above, the \opf Problem \eqref{eq::ACOPF} can be stated in affinely coupled separable form
\begin{subequations} \label{eq:OPF}
	\begin{align} 
	&\min_{x} && \sum_{i\in \mathcal{R}} f_i(x_i) \label{eq:SepProbObj} \\
	&\;\;\text{s.t.}&&\sum_{i\in \mathcal{R}}A_i x_i=0\;\mid\lambda,\label{eq:consCnstr}\\[0.1cm]
	&&  &h_{i}(x_i) = 0 \quad \;\, \mid \kappa_i  &\forall i \in \mathcal{R}, \label{eq:SepProbEqcnstr}\\[0.1cm] 
	&&  &\underline{x}_i \leq x_i \leq  \overline{x}_i\;\, \mid\eta_i  &\forall i \in \mathcal{R}.\label{eq:SepProbIneqcnstr}
	\end{align}
\end{subequations}
Here, $x=[x_1^\top, \dots, x_R^\top]^\top \in \mathbb{R}^{n_x}$ stacks the local decision vectors $x_i$, and $\lambda,\;\kappa_i,\;\eta_i$ denote the dual variables (multipliers) of the respective constraints.
At all auxiliary bus pairs \mbox{$(k,l) \in \mathcal{A}$}, we enforce consensus in the physical values
\begin{equation}
\label{eq::couplings}
\theta_k = \theta_l,\;\;\; v_k = v_l,\;\;\; p_k = -p_l,\;\;\; q_k = -q_l,
\end{equation}
which leads to the affine consensus constraint \eqref{eq:consCnstr}. 
The box constraints~\eqref{eq:SepProbIneqcnstr} collect local bounds on active/reactive power injections and voltage magnitudes for all regions.

\section{ALADIN-based Distributed OPF}
\label{sec:ALADIN}
We describe a variant of \aladin for solving Problem \eqref{eq:OPF} in distributed fashion, cf. Algorithm~\ref{alg:ALADIN}.
\aladin consists of five steps:
\begin{algorithm}[hbtp!]
	\small
	\caption{\aladin-based Distributed \opf}
	\textbf{Initialization:} Initial guess $(z^0,\lambda^0)$, choose $\Sigma_i,\rho^0,\mu^0,\epsilon$. \\
	\textbf{Repeat:}
	\begin{enumerate}
		\item \comm{\textit{Parallelizable Step:} \label{step:parStep}
		Solve for each $i \in \mathcal{R}$
		\begin{align}
		\begin{split}
		\label{eq::denlp}
		\underset{x_i\in[\underline x_i, \overline x_i]}{\min}& f_i(x_i) + (\lambda^k)^\top A_i x_i + \hspace{-0.1em} \frac{\rho^k}{2}\left\|x_i-z_i^k\right\|_{\Sigma_i}^2\\
		\text{s.t.}\quad& h_i(x_i) = 0 \quad \mid \kappa_i^k \quad 
		\end{split}	
		\end{align}
		either
		\begin{enumerate}[label=\roman*)]
		\item exactly, obtaining $x_i^{\star{}k}$ and assigning $x_i^k:=x_i^{\star{}k}$ \label{step:exact}; or
		\item approximately, obtaining $\bar x_i^k$ 
		and 	 assigning $x_i^k:=\bar x_i^{k}$.  \label{step:approx}
	    \end{enumerate}}
		
		\item \textit{Termination Criterion:} If 
		\begin{equation}
		\label{eq::stop}
		\left\|\sum_{i\in \mathcal{R}}A_ix^k_i\right\|\leq \epsilon \text{ and } \left\| x^k - z^k \right \|\leq \epsilon\;,
		\end{equation}
		return $x^\star = x^k$.
		
		\item \textit{Sensitivity Evaluations:} Compute and communicate local gradients~$g_i^k=\nabla f_i(x_i^k)$,
		Hessian approximations~{$B_i^k \approx \nabla^2\{f_i(x_i^k)+\kappa_i^\top h_i(x_i^k)\}$} and constraint Jacobians $C^k_i = \nabla h_i(x^k_i)$. 
		
		\item \textit{Consensus Step:} Solve the coordination \qp 
	    \begin{align}
		\notag
	 &\underset{\Delta x,s}{\min}\;\;\sum_{i\in \mathcal{R}}\left\{\frac{1}{2}\Delta x_i^\top B^k_i\Delta x_i + {g_i^k}^\top \Delta x_i\right\}     + (\lambda^k)^\top s + \frac{\mu^k}{2}\|s\|^2_2  \\ 
        &\begin{aligned}\label{eq::conqp} 
		\text{s.t.}\;                                  &\quad \sum_{i\in \mathcal{R}}A_i(x^k_i+\Delta x_i) =  s     &&|\, \lambda^\mathrm{QP},\\[0.2cm]
		                                                  &\quad C^k_i \Delta x_i = 0                                    &&\forall i\in \mathcal{R},\\[0.2cm]
		                                                  &\quad  (\Delta x_i)_j  = 0  \quad j\in\mathbb{A}^k_i          &&\forall i\in \mathcal{R},
		\end{aligned}
		\end{align}
		\comm{obtaining $\Delta x^k$ and $\lambda^{\text{QP}}$ as the solution of \eqref{eq::conqp}}.

		\item \textit{Line Search:} Update primal and dual variables by
		\begin{eqnarray}\notag
		z^{k+1}&\leftarrow&z^k + \alpha^k_1(x^k-z^k) + \alpha_2^k\Delta x^k\;,\\[0.2cm]\notag
		\lambda^{k+1}&\leftarrow&\lambda^k + \alpha^k_3 (\lambda^\mathrm{QP}-\lambda^k),
		\end{eqnarray}
		with $\alpha^k_1,\alpha^k_2,\alpha^k_3$ from~\cite{Houska2016}. If full step 
		is accepted, i.e. $
		\alpha_1^k=\alpha_2^k=\alpha_3^k=1,
		$
		update $\rho^k$ and $\mu^k$ by
		\begin{align}
		\rho^{k+1}\;(\mu^{k+1}) =
		\begin{cases}
		r_\rho \rho^k\;(r_\mu \mu^k)   &\text{if} \; \rho^k < \bar \rho\;\; (\mu^k < \bar \mu)\\ \nonumber
		\rho^k\;(\mu^k)  &\text{otherwise} 
		\end{cases} .
		\end{align}
	\end{enumerate}
		\label{alg:ALADIN}
\end{algorithm}
\begin{enumerate}
\item \comm{Solve the decoupled \nlps~\eqref{eq::denlp} in parallel either
\begin{enumerate}[label=\roman*)]
	\item exactly, which yields (modulo technical assumptions) global convergence guarantees and fast local convergence, cf. Theorem \ref{ConvTheoremALADIN}; or
	\item approximately, preserving fast local convergence, cf. Theorem \ref{the::localcon}.
\end{enumerate}}
\item If the solution satisfies the termination criterion~\eqref{eq::stop} ($\epsilon$ is choosen by the user), terminate with the solution from the local \nlps, $x^\star=x^k$. 
\item If not, compute gradients~$g^k_i$, Hessian approximations $B_i^k$ and constraint Jacobians $C^k_i$. Note that these computations are  fully parallelizable.\footnote{In case of derivative--based solvers, these sensitivities can be obtained from the local solvers for \eqref{eq::denlp} avoiding explicit evaluation.}$^,$\footnote{\aladin requires the Hessian approximations $B_i^k$ to be positive definite to ensure convergence \cite{Houska2016}. To ensure positive definiteness, we toggle the sign of all negative eigenvalues of all $B_i^k$s and add a small positive constant to all zero eigenvalues.}
\item Construct the consensus \qp~\eqref{eq::conqp} based on local sensitivities and the active sets
\[
\mathbb A_i^k  =\{\;j\;\mid  (x^k_i)_j = \underline{x}_i \text{ or } \overline{x}_i\;\}\,
\]
detected by the local \nlps.
Note that there are no inequality constraints in the \qp~\eqref{eq::conqp}. Thus, solving this problem is equivalent to solving a linear system  of equations yielding a computationally cheap numerical operation~\cite{Nocedal2006}.
\item Apply the globalization strategy proposed in~\cite{Houska2016} to update $x^k$ and $\lambda^k$. In practice, full steps are often accepted and the line search can be omitted. Finally, update the parameters $\rho^k$ and $\mu^k$. 
\end{enumerate}
Algorithm~\ref{alg:ALADIN} provides technical details to \aladin. 

We remark that instead of computing exact Hessians  in Step~3), one can also use approximation techniques for $B_i^k$ based on previous gradient evaluations. Here, we use the blockwise and damped Broyden-Fletcher-Goldfarb-Shanno (\bfgs) update. In contrast to standard \bfgs, the damped version ensures positive definiteness of the $B_i^k$s to preserve the convergence properties of \aladin, cf. \cite{Houska2016,Nocedal2006}. The \bfgs update formula is given by 
\begin{equation} \label{eq:BFGS}
B_i^{k+1} = B_i^k - \frac{B^k_i s_i^k s_i^{k\top} B_i^k}{s_i^{k\top} B_i^k s_i^k} + \frac{r_i^k r_i^{k\top}}{s_i^{k\top} r_i^k}
\end{equation}
with $s_i^k=x_i^{k+1}-x_i^k$ and $r_i^k=\theta^k(g_i^{k+1}(\lambda^{k+1})- g_i^k(\lambda^{k+1})) + (1-\theta^k)B_i^ks_i^k$, where $g_i^k(\lambda)=\rho^k(z_i^k-x_i^k)-A_i^\top\lambda$ are the gradients of the Lagrangians \cite{Nocedal2006}.
The damping parameter $\theta^k$ is computed by the update rule given in \cite[p. 537]{Nocedal2006}.
Notice that  \bfgs reduces the need for communication within \aladin: instead of the full Hessian matrix, it suffices to communicate the gradients of the Lagrangians, and then update $B_i^k$ in the coordination Step 4).

\comm{In contrast to \admm, \aladin provides convergence guarantees for non-convex optimization problems such as \ac-\opf.
	As we recall next, in case of applying Step \ref{step:parStep}) \ref{step:exact} of \aladin global convergence (i.e. convergence with arbitrary initialization) is achieved.}

\begin{assumption}[Problem data and \aladin parameters]
	\label{ass:GlobalConvergence}
	\comm{
	\begin{enumerate}[label=\roman*)]
		\item \label{item:ProofAss1} Problem~\eqref{eq:OPF} has a compact feasible set. Moreover, linear independence constraint qualification, strict complementarity	conditions, as well as the second-order sufficient condition are satisfied at all local minimizers.
		\item \label{item:ProofAss3} For all $i\in\mathcal{R}$, the functions $f_i$ and $h_i$ are twice Lipschitz-continuously differentiable  on the local feasible sets ${\mathcal{F}_i=\{ x_i \; | \; h_i(x_i) = 0,\; \underline{x} \leq x \leq \bar{x}\}}$.
		\item \label{item:ProofAss4} The matrices $\Sigma_i$ from \eqref{eq::denlp} are positive definite.
		\item \label{item:ProofAss5} The parameters $\rho$ and $\mu$ are sufficiently large and the line search parameters are adjusted by the globalization strategy stated in \cite{Houska2016}.
		\hfill $\blacksquare$
	\end{enumerate}
}
\end{assumption}
\comm{Note that assumptions \ref{item:ProofAss1}-\ref{item:ProofAss3} are not very restrictive standard assumptions from optimization theory and often satisfied in practice. For example see \cite{Hauswirth18a} for the discussion of linear independence constraint qualifications in \opf.
	Assumptions \ref{item:ProofAss4} and \ref{item:ProofAss5} can be satisfied by choosing appropriate parameters/matrices.}
\begin{theorem}[Global convergence of \aladin] ~\\ \label{ConvTheoremALADIN} 
	\comm{If Assumption~\ref{ass:GlobalConvergence} holds, then Algorithm \ref{alg:ALADIN} executed with Step \ref{step:parStep}) \ref{step:exact} terminates for any user-specified tolerance $\epsilon > 0$ after a finite number of iterations. 
		$\hfill \blacksquare$}
\end{theorem}
\comm{For the details of the proof we refer to \cite[Thm. 2]{Houska2016}. From  Step 2) it follows that upon termination \aladin returns a solution satisfying $\left\|\sum_{i\in \mathcal{R}}A_ix^k_i\right\|\leq \epsilon$.   }
%
Regarding the convergence rate, quadratic (respectively superlinear for BFGS variants) convergence is shown for \aladin in case the \nlps~\eqref{eq::denlp} are solved to optimality \cite{Houska2016}.
	However,  in practice, due to finite precision arithmetics, numerical solvers do not return truly exact solutions.
		Next we extend the results from~\cite{Houska2016} to cover this. 
\begin{assumption}[Accuracy of local \nlp solutions]
	\label{ass:TechnicalConditions}~\\
	For all iterations $k\in \mathbb{N}$, the following holds:
	\comm{
	\begin{enumerate}[label=\roman*)]
	\item \label{ass::inesol} The approximate solution $\bar x^k$ satisfies 
	\begin{equation}
	\label{eq::inexact}
	\|\overline{x}^k - x^{k}\| \leq \zeta_1\|z^k - x^{k} \|
	\end{equation}
	with constant $\zeta_1>0$.
	\item \label{ass::rho}
	The penalty parameter $\rho^k>0$ in Problem~\eqref{eq::denlp} satisfies 
	\begin{equation}
	\nabla^2 \{f_i(x_i^k) + {\kappa_i^k}^\top h_i(x_i^k) \} + \rho^k \Sigma_i \succ 0 
	\end{equation}
	for all $i=1,\dots,R$. 	\hfill $\blacksquare$
	\end{enumerate}}
\end{assumption}	
\comm{
	Note that item~\ref{ass::inesol} of   Assumption~\ref{ass:TechnicalConditions} can be satisfied e.g. by choosing $\zeta_1=1$ and $\bar x_k = z_k$. In this case, \aladin is equivalent to \sqp as no local steps are computed. On the other hand, if we solve the local \nlp{}s exactly, we obtain \aladin in its pure form, cf. \cite{Houska2016}. From this perspective, approximating a minimizer of the \nlps yields an algorithm in-between  \sqp and (exact) \aladin. 
	Item~\ref{ass::rho} of Assumption~\ref{ass:TechnicalConditions} is not very restrictive as it can be satisfied by choosing $\rho^k$ sufficiently large.
However, note that in case of minimizer approximations the global convergence Theorem~\ref{ConvTheoremALADIN} fails to hold.}
\comm{
\begin{theorem}[Local quadratic convergence of \aladin]~\\ 
	\label{the::localcon}
	Let Assumption~\ref{ass:GlobalConvergence} hold and  
	let  $\rho^k>0$ and $\bar x^k$ satisfy Assumption~{\ref{ass:TechnicalConditions}}.
	Suppose that Algorithm~\ref{alg:ALADIN} executed with Step \ref{step:parStep}) \ref{step:approx} 
	\begin{itemize}
	\item is initialized with $(x^0,\lambda^0)$ close to $(x^\star,\lambda^\star)$;
	\item that Step~3) computes exact sensitivities $B_i^k=\nabla \{f_i(x_i^k) + {\kappa_i^k}^\top h_i(x_i^k) \}$ and $C_i^k = \nabla h_i(x_i^k)$;
	\item and additionally, the update of $\mu^k$ in Step~5) satisfies 
	\begin{equation}
	\label{eq::mu}
	\frac{1}{\mu^k}\leq \mathbf{O}(\|\overline{x}^k - x^\star\|)\;.
	\end{equation}
	\end{itemize}
	Then 
	the iterates $(z^k,\lambda^k)$ converge  locally to $(x^\star,\lambda^\star)$ at a quadratic rate.
	\hfill $\blacksquare$
\end{theorem}}
The proof is given in Appendix~\ref{sec:LocConvProof}.
We remark that \eqref{eq::mu} can be satisfied by choosing an appropriate update rule for $\mu^k$.

\comm{\begin{remark}[Superlinear convergence for  \aladin-\bfgs]
		\label{rema:Convergence}
		With minor modifications, the proof of Theorem \ref{the::localcon} can be extended to cover \aladin-\bfgs. In this case, one obtains superlinear convergence rate provided that the Hessians and Jacobians converge to their optimal counterparts, i.e. $B_i^k \rightarrow \nabla^2\{f_i(x_i^\star)+\kappa_i^\top h_i(x_i^\star)\}$ and $C_i^k\rightarrow \nabla h_i(x_i^\star)$. 
\end{remark}}

\section{Numerical Results}
\label{sec:results}
The presentation of our results is divided into three parts: We being by showing considerable performance differences of \aladin and \admm for a motivating 5-bus example depicted in Fig.~\ref{fig:grid}. Moreover, we illustrate that \aladin performs well for larger grids (30-bus, 57-bus) when inexact Hessians are used. 
Finally, we apply \aladin to the 118 and 300-bus test cases, and compare our results to variants of \admm published in the literature.

All units are given in p.u. for a base power of 100\,MVA.
In all cases, we initialize with voltage magnitudes of $1$ p.u.; all other values are set to zero initially (flat start).
The dual variables $\lambda$ are initialized with zero. 
We compare \aladin and \admm in terms of number of iterations,  as well as computation times and communication effort.

Our implementation uses the CasADi toolbox \cite{Andersson2013b} running with MATLAB R2016a and IPOPT \cite{Waechter2006} as solver for the local \nlps. 
The ``true'' minimizers $x^\star$ are obtained by solving problem \eqref{eq:OPF}  with IPOPT centrally.

\subsection{5-bus System with Line Limits}

\begin{figure*}[h]
	\centering
	\adjustbox{trim={.07\width} {.0\height} {0.08\width} {.0\height},clip}
	{\includegraphics[width=1.18\linewidth]{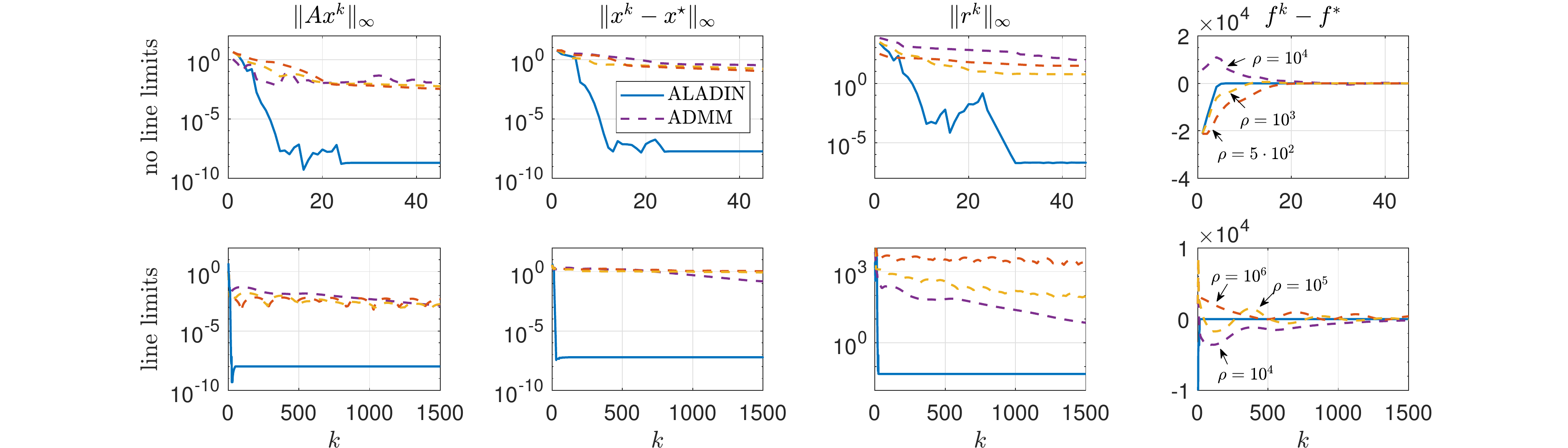}}
	\caption{Convergence of \aladin (solid) and \admm (dashed) for the 5-bus system with and without considering line limits.}
	\label{fig:Lcstr}
\end{figure*}

\begin{figure}[h]
	\centering
	\adjustbox{trim={.0\width} {.51\height} {0.08\width} {.05\height},clip}
	{\includegraphics[width=1\linewidth]{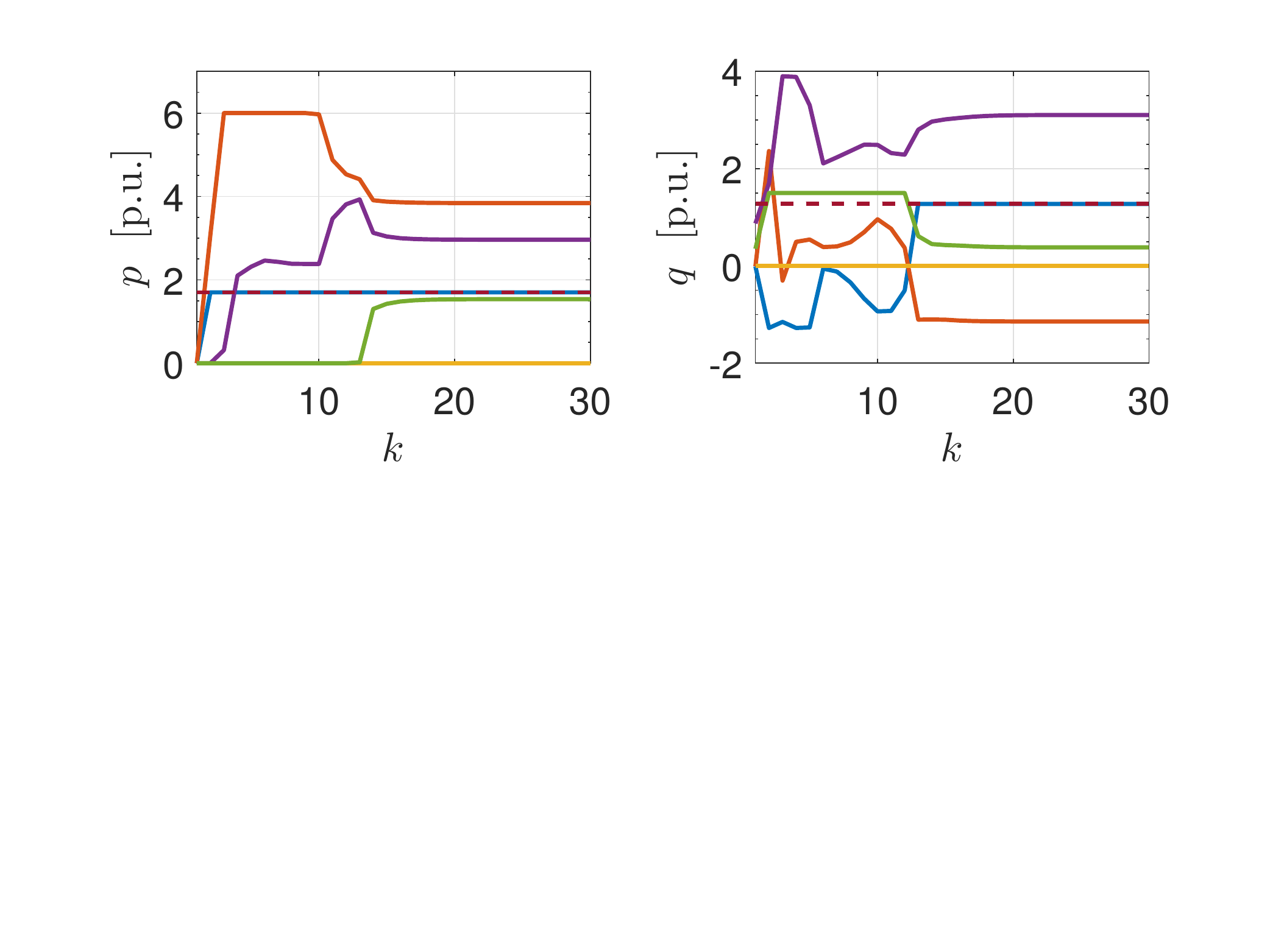}}
	\caption{Power injections and selected limits (dashed) over the iteration index $k$ for the 5 bus system considering line limits.}
	\label{fig:physVals}
\end{figure}

Consider the 5-bus case with line limits as shown in Fig. \ref{fig:grid} in order to compare \aladin and \admm.
We  partition the grid into three regions such that it is expected to be difficult for both distributed optimization algorithms.
Specifically, there is a generation center in the west with cheap generators and no loads, which means that large amounts of power have to be transferred to the load centers located in the east.
Moreover, line limits between these regions are active (between buses (1,\,2) and (4,\,5)).

Many works using \admm for \opf do not consider line limits \cite{Erseghe2014a,Erseghe2015} as they add additional nonlinear inequality constraints to the problems.
The recently published work \cite{Guo2017} is one of the few that explicitly considers line limits.
Here, they are considered as limits on the magnitude of the apparent power\footnote{\label{footn:lineLimits}\comm{In Algorithm \ref{alg:ALADIN}, these limits \eqref{eq:lineLimits} are considered  by introducing additional decision variables $s_{kl}^l$ constrained by $s_{kl}^l = p_{kl}^2 + q_{kl}^2$ and $s_{kl}^l\leq |\bar s_{kl}|$ respectively.}}
\begin{equation} \label{eq:lineLimits}
p_{kl}^2+q_{kl}^2 \leq |\bar s_{kl}|^2,
\end{equation}
where
\begin{align*}
p_{kl}&= - v_k^2G_{kl} + v_{k}v_{l}(G_{kl}\cos(\theta_{k}-\theta_{l})+B_{kl}\sin(\theta_{k}-\theta_{l})), \\
q_{kl}&=\phantom{-} v_k^2B_{kl} - v_{k}v_{l}(B_{kl}\cos(\theta_{k}-\theta_{l})-G_{kl}\sin(\theta_{k}-\theta_{l})).
\end{align*}
Due to the non-convexity of these constraints they are difficult to handle; especially when they are located at lines connecting regions. 

\comm{Applying \aladin to the 5-bus system requires to select tuning parameters $\rho^k$ and $\mu^k$. Values for these parameters are determined by parameter sweeps for each grid aiming for fast convergence. The results are shown in Table \ref{tab::prameters}. 
To obtain a similar scaling, the weighting matrices $\Sigma_i$ are chosen such that each diagonal entry is inversely proportional to its corresponding decision variable range.
Therefore, entries corresponding the power injections are chosen to 1; entries corresponding to voltage magnitudes and voltage angles are chosen to 100.}

\comm{Fig. \ref{fig:physVals} shows active/reactive power injections and line flows $s_{kl}$  over the iteration index $k$ computed by \aladin for the 5-bus system with line limits. \aladin reaches the final (and optimal) values in around 15 iterations and satisfies active/reactive power limits (dashed). }

\comm{
In the following, we compare the performance of \admm and \aladin in terms of the following convergence criteria:}
\comm{
\begin{itemize}
	\item The consensus violation $\|Ax^k\|_\infty$ with $A = [ A_1,\dots,A_R ]$ indicates the maximum mismatch of voltages/powers at auxiliary buses.
	\item The distance to the minimizer $\|x^k-x^\star\|_\infty$ is the maximum distance of the current power/voltage iterates to its optimal value, where  $x^\star$ is the ``true" minimizer obtained by solving \eqref{eq:OPF} in centralized fashion.
	\item The inf-norm $\|r^k \|_{\infty}$ of the dual residual
	\begin{equation*}
	r^k = \sum_{i\in\mathcal{R}}\left\{\nabla f_i(x_i^k) + A_i^\top \lambda^k +\nabla h_i(x_i^k) \kappa_i^k + \eta_i^k\right\}
	\end{equation*}
	measures violation of the first-order optimality conditions.
	\item The suboptimality gap $f(x^k)-f(x^\star)$.
\end{itemize}
We remark that for \aladin \emph{and} \admm the generated iterates always satisfy the nonlinear equality/inequality constraints \eqref{eq:SepProbEqcnstr} and \eqref{eq:SepProbIneqcnstr} as they are explicitly considered in the local \nlps \eqref{eq::denlp}. Hence, it is sufficient to show the consensus violation to ensure satisfaction of the power flow equations and limits (feasibility). Optimality is indicated by the remaining indicators suboptimality and distance to the minimizer. }

Fig. \ref{fig:Lcstr} shows how the convergence criteria for \aladin and \admm when applied to the 5-bus system in two settings:
In the first setting line limits are neglected, while in the second setting there are apparent power limits at the lines (1,\,2) and (4,\,5) of 240\,MVA and 180\,MVA respectively.
To enable a fair comparison, the penalty parameters $\rho$ for \admm are chosen based on parameter sweeps aiming for fast convergence.

Without line limits, \aladin converges around 3-5 times faster than \admm.
However, with slight abuse of optimality and consensus, applicable solutions can be obtained via \admm in around 50 iterations assuming that underlying frequency controllers account for the remaining power mismatch.

In case of active line limits, \aladin takes around 30 iterations to converge to the exact solution whereas \admm requires around 1500 iterations to reach the medium level of accuracy as above.  
Observe that \aladin seems to converge at quadratic rate, which is in line with Theorem~\ref{the::localcon}.
For \admm we expect at most a linear convergence rate (as this is the rate achieved by \admm for convex problems) which coincides with the seemingly slow convergence especially in case of binding line limits, cf. Fig. \ref{fig:Lcstr}.

\subsection{30-bus and 57-bus  with Inexact Hessians}

\begin{figure*}[h]
	\centering
	\adjustbox{trim={.06\width} {.5\height} {0.08\width} {.03\height},clip}
	{\includegraphics[width=1.15\textwidth]{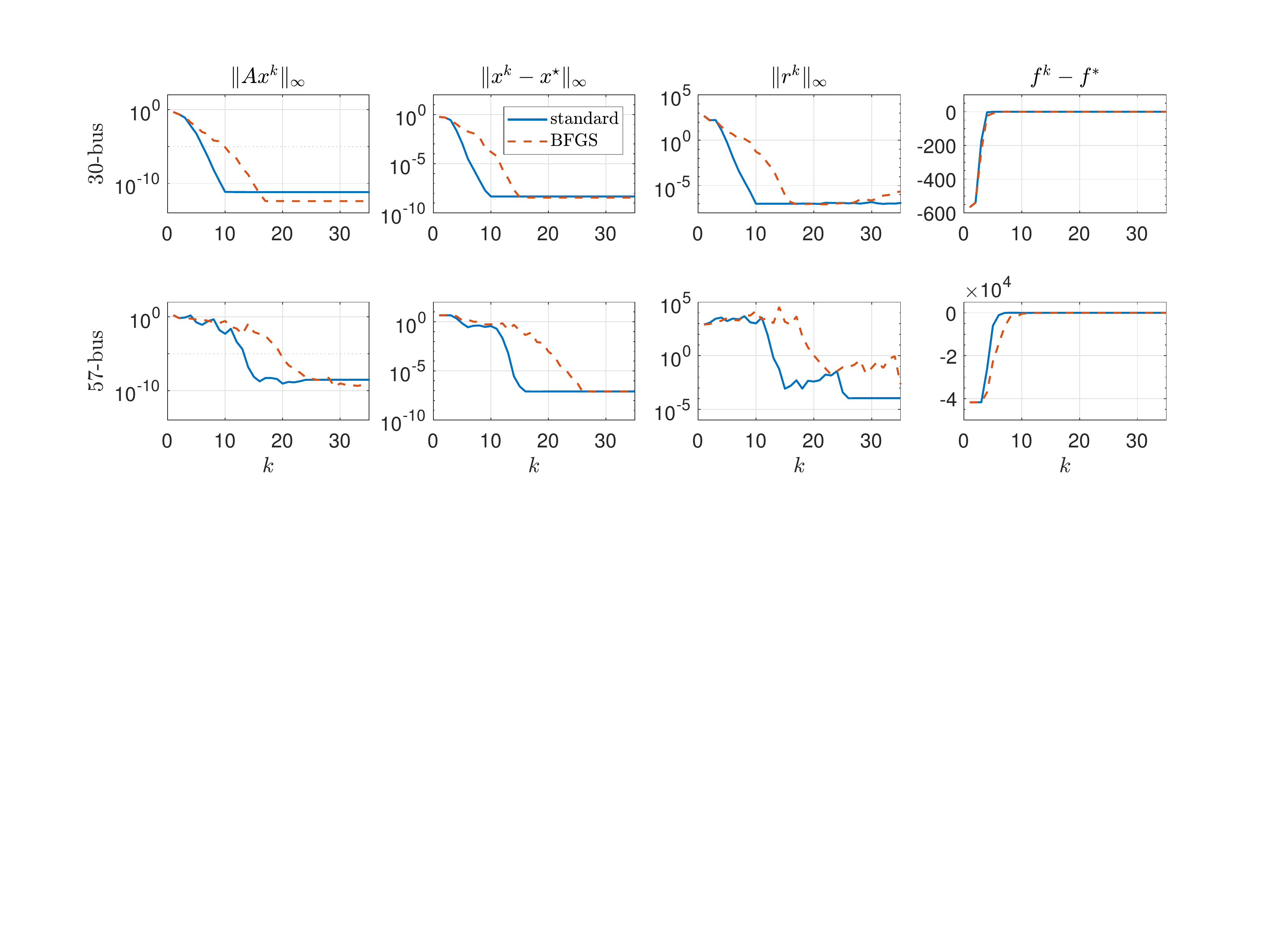}}
	\caption{Convergence behavior of \aladin for the \ieee 30-bus and 57-bus test cases using exact Hessians (solid) and inexact Hessians (dashed, here \bfgs).}
	\label{fig:BFGS}
\end{figure*}

Next, we compare the performance of \aladin with exact Hessians to \aladin with inexact Hessian for larger grids.
Specifically, we use approximations based on the \bfgs formula~\eqref{eq:BFGS}.
Inexact Hessians reduce the per-step communication effort, which is advantageous.
The employed grid partitioning for the considered \textsc{ieee} 30- and 57-bus test cases are taken from \cite{Erseghe2014a}, and listed in Table~\ref{tab::partitioning} in the Appendix for self-containment.

To foster numerical convergence we add a quadratic regularization for the reactive power injection to the local objective functions 
\begin{equation*}
\tilde f_i(x_i) = f_i(x_i) + \gamma \sum_{k \in \mathcal{N}_i} q_k^2
\end{equation*}
with $\gamma$ non-negative in the rest of the paper. This regularization follows the technical motivation to keep reactive power injections small.
We choose $\gamma = 10\, \frac{\$}{hr\cdot(p.u.)^2}$ which is around 10\,\% of the quadratic coefficient of the active power injections $c_{1,k}$.

Fig. \ref{fig:BFGS} depicts the convergence behavior of \aladin with exact and inexact Hessians.\footnote{The centralized minimizer $x^\star$ is computed here including the regularization into the objective of \eqref{eq:OPF}.}
For both cases \aladin converges in less than 40 iterations to high accuracy (at least 10\textsuperscript{\,-\,4} for all convergence criteria).
Furthermore,  Fig.~\ref{fig:BFGS} shows that \aladin with inexact Hessians needs just slightly more iterations compared with \aladin using exact Hessians.
One can observe that the convergence rate for \aladin using inexact Hessians seems to be faster than linear.
This observation is consistent with Theorem \ref{the::localcon}.

\subsection{118-bus and 300-bus \aladin vs. \admm}
\label{sec:118and300}
For the \textsc{ieee} 118-bus and 300-bus test cases, we compare \aladin with exact Hessians to \admm results documented in the literature \cite{Erseghe2014a,Erseghe2015} 
\comm{supposing the authors thereof chose the parameters and their update rules optimally to facilitate fast convergence.}
We also adopt the grid partitioning from \cite{Erseghe2014a} for the 118-bus case.
Unfortunately, the partitioning for the 300-bus case is not given in \cite{Erseghe2014a}. Hence we choose the partitioning given in Table~\ref{tab::partitioning} in the Appendix. 

\begin{figure*}[h]
	\centering
	\adjustbox{trim={.06\width} {.04\height} {0.08\width} {.48\height},clip}
	{\includegraphics[width=1.15\textwidth]{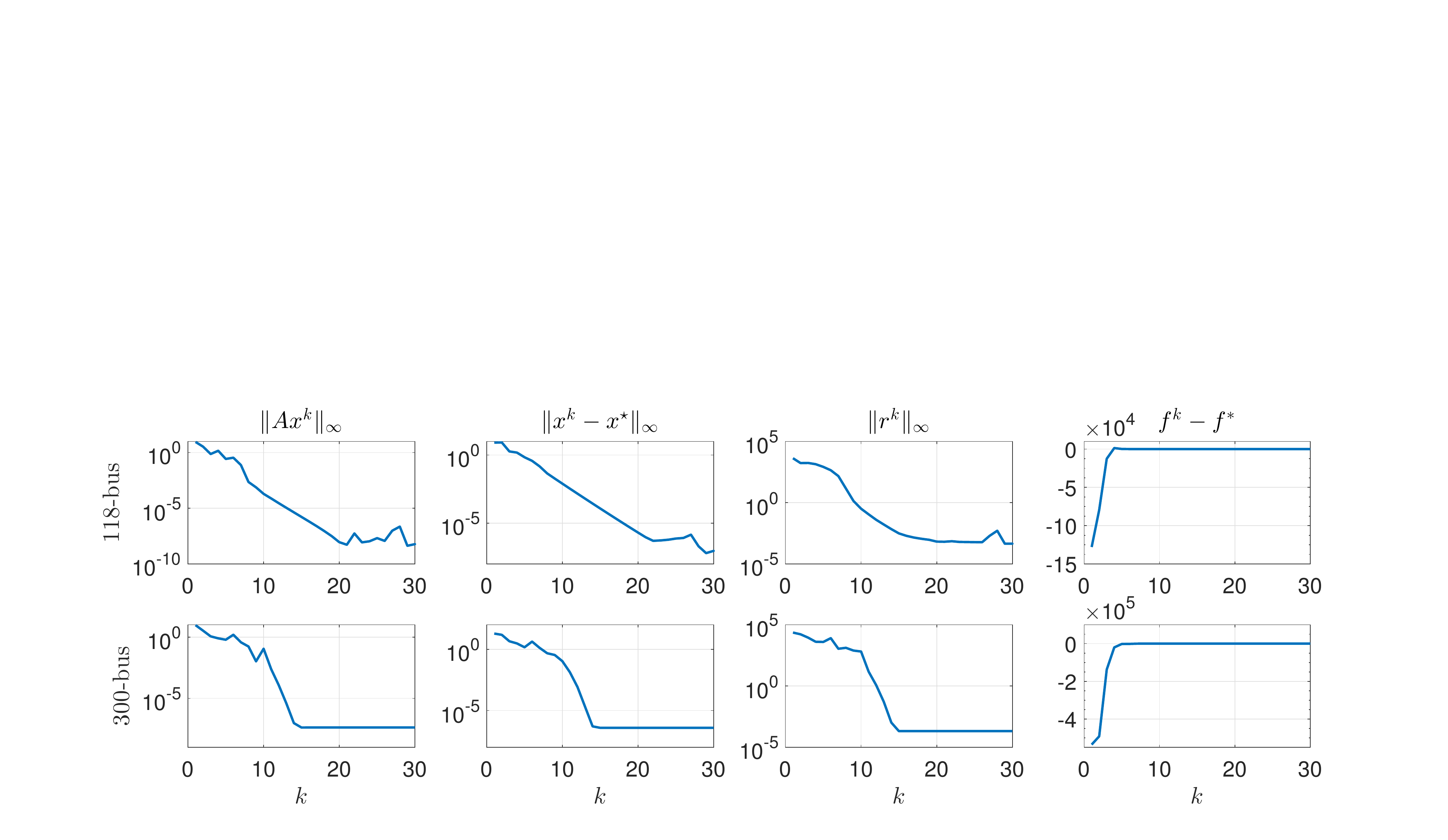}}
	\caption{\comm{Convergence behavior of \aladin for the \ieee 118 and 300-bus test cases.}}
	\label{fig:allADMMlit}
\end{figure*}

Using \aladin, we obtain the numerical results for the 118-bus and 300-bus system shown in Fig. \ref{fig:allADMMlit}.
In either case, \aladin shows fast convergence to a high level of accuracy for all convergence criteria.
In \cite{Erseghe2014a,Erseghe2015} the main convergence criterion is taken to be the infinity norm of the primal gap $\|Ax^k\|_{\infty}< \epsilon$.
Adopting this criterion allows a direct comparison between \textsc{aladin} and \admm results from \cite{Erseghe2014a,Erseghe2015}; for $\epsilon = 10^{-4}$ the results are summarized in Table~\ref{tab::comparison}.\footnote{We remark that primal feasibility does not ensure convergence to a minimizer, cf.  \cite[Sec. 3.3.1]{Boyd2011} for an \admm-specific discussion. This lack of optimality guarantees can be observed in the numerical results in \cite{Erseghe2014a,Erseghe2015}. 
However, in practice small optimality gaps are often accepted. Nontheless one has to bear in mind that using $\| Ax^k\| \leq \epsilon$ does not imply convergence of the reactive power injections to the optimal ones since the sensitivity of the objective function with respect to the reactive power is much smaller than the sensitivity to active power. This can be verified by comparing the dual variables for active and reactive power injections, cf. \cite[Chap. 3.2.3]{Bertsekas1999}.} 
\aladin converges around one order of magnitude faster while much higher accuracies in terms of the optimality gap and dual residual are obtained.

\begin{table}
	\centering
	\caption{Comparison of tuned \admm from \cite{Erseghe2014a,Erseghe2015} and \aladin employing $\|Ax^k\|_\infty\leq 10^{-4}$ as convergence criterion only. 
	 }
	\renewcommand{\arraystretch}{1.7}
	\begin{tabular}{c|cccc}
		\toprule
		          & \multicolumn{2}{c}{\admm} &     \multicolumn{2}{c}{\aladin}       \\ \midrule
		Test Case & \#Iter &                  $\left |\frac{f(x)-f(x^\star)}{f(x^\star)}\right|$                   & \#Iter &         $\left |\frac{f(x)-f(x^\star)}{f(x^\star)}\right |$           \\
		   30     &   110    &              0.14\phantom{0} \,\%              &   6    & 4.50 $\cdot$ 10\textsuperscript{\,-\,3}\, \%   \\
		   57     &   144    &                   0.002\,\%                    &   13   & 2.91 $\cdot$ 10\textsuperscript{\,-\,4}\, \%  \\
		   118    &   186    &              0.25\phantom{0}\,\%               &   11   & 3.86 $\cdot$ 10\textsuperscript{\,-\,5}\, \%  \\
		   300    &   216    &              0.23\phantom{0}\,\%               &   26   & 4.26 $\cdot$ 10\textsuperscript{\,-\,5}\, \%  \\ \bottomrule
	\end{tabular}
	\label{tab::comparison}
\end{table}

\begin{table}
	\caption{Comparison of algorithmic properties. The number of worst case forward (backward) communications  in terms of floats is denoted by $\nFW$ ($\nBW$).}
	\centering
	\renewcommand{\arraystretch}{1.7}
	\begin{tabular}{p{1.5cm}ccc}
		\toprule
		& \admm  & \aladin & \aladin-\bfgs \\
		\midrule 
		Convergence guarantee	& no & yes & yes \\ 
		Convergence  rate  & (linear) & quadratic& superlinear \\
		$\nFW $  &$ \displaystyle \sum_{i\in \mathcal{R}} n_i$& $  \displaystyle \sum_{i\in \mathcal{R}}  \frac{n_{i} (2n_{i}+3)}{2}$   & $ \displaystyle \sum_{i\in \mathcal{R}}  \frac{ n_i(n_{i}+4)}{2}$\\
		$\nBW $& $\displaystyle \sum_{i\in \mathcal{R}}  n_i$ & $ \displaystyle \sum_{i\in \mathcal{R}}  2n_i$ & $ \displaystyle \sum_{i\in \mathcal{R}}  2n_i$ \\
		\bottomrule		
	\end{tabular}
	\label{tab::comparisonAll}
\end{table}

\section{Discussion---\aladin vs. \admm} \label{sec::ALvsADM}

Our numerical results from Section~\ref{sec:results} using \aladin seem promising.
However, compared with \admm there is an increased per-step communication effort when employing \aladin.
Thus, we discuss how to trade-off convergence behavior and convergence guarantees versus per-step communication effort.

\begin{table*}[t]
	\caption{Worst case computation times (in s) and worst case forward communication effort (in floats).	\label{tab::Comm&timing}}
	\scriptsize
	\centering
	\renewcommand{\arraystretch}{1.7}
	\begin{tabular}{p{0.5cm}|c|cccc|ccccc|ccccc}
		\toprule 
		&  & \multicolumn{4}{c}{\admm }    & \multicolumn{5}{c}{\aladin }& \multicolumn{5}{c}{\aladin-\bfgs}   \\ 
		Test Case  & $\tNLP$ & $\tWC$ & $\nFWeff$ &$\nFW$ & $\nFWeff \cdot \text{\#Iter}$  &$ \tWC $&   $\tQP$ & $ \nFWeff$  &  $ \nFW$&$ \nFWeff \cdot \text{\#Iter}$ & $ \tWC $ & $\tQP$ & $ \nFWeff$  & $\nFW$ & $\nFWeff \cdot \text{\#Iter}$ \\ 
		\midrule 
		30 &  0.03  & 3.30 & 32&184 &3,520 & 0.2 &  0.004 & 2,213 & 8,916 &13,278& 0.28 & 0.005 & 1,012		& 4,688 & 8.096\\		
		57&   0.04  & 5.76 & 96 &420 &13,824 &0.66&  0.011 & 5,527 & 23,814 &71,851		&0.98 &0.012 & 2,225		& 12,432& 42.275\\	
		118&  0.05   & 9.30 & 52&576& 9,672 &0.76&  0.019 & 14,412 & 86,208 &158,532		&	-	& - & - & - & -\\	
		300&    0.15 & 32.40 & 244&1,688 &52,704 & 14&  0.39 & 129,664 & 955,652 &3,371,264		& -	&  - & - & - & -\\	
		\bottomrule		 	
	\end{tabular}
\normalsize
\end{table*}
\subsubsection{Convergence Properties}
\admm and \aladin exhibit differences in convergence guarantees.
In case of \admm, a linear convergence rate can be achieved for strictly convex problems under rather mild assumptions like Lipschitz continuity of the gradient and regularity assumptions on the affine constraints \cite{Deng2016}.
In case of convex problems, sublinear convergence is achieved \cite{Deng2016}.
For the non-convex case, convergence can only be guaranteed for special problem classes, where---to the best of the authors' knowledge---it is not clear whether \ac-\opf belongs to them \cite{Hong2016}.
However, this does not mean that \admm does not work for non-convex \opf.
Yet, one has to be aware that \admm does not necessarily converge to a local minimizer, or converge at all.
Nevertheless, \admm  works well in practice but often shows slow practical convergence rates, especially if high accuracies are needed \cite{Boyd2011}.
This is in accordance with the simulation results from Section~\ref{sec::ALvsADM} and the result of \cite{Erseghe2014a,Erseghe2015}.

\comm{As shown in Section \ref{sec:ALADIN}, convergence for \aladin can be guaranteed without relying on a convexity assumption of the objective or the constraints (Theorem \ref{ConvTheoremALADIN}).
Only mild assumptions on the penalty parameter as well as Lipschitz continuity are required.
In case of  Hessian approximation via \bfgs updates, superlinear convergence can be achieved while in case of exact Hessians quadratic convergence is guaranteed (Theorem \ref{the::localcon}).
This comes at the cost of an increased per-step communication, and the need for a central coordinating entity that has to solve the coupling \qp. Furthermore \aladin requires a communication link to this coordinator.}
\subsubsection{Worst Case Communication Effort}

\comm{The main conceptual difference between \aladin and \admm is that \aladin uses second-order information whereas \admm only communicates local primal solutions.
More specifically, \aladin relies on communicating local sensitivities and the active sets, i.e.
$
g_i,\;B_i, \;C_i,\;\mathbb{A}_i\,.
$
for all regions $i\in\mathcal{R}$, cf. Step 3) of Algorithm \ref{alg:ALADIN}.
The gradients $g_i$ are of dimension $n_i$, the (symmetric) Hessians $B_i$ of dimension $n_i \times n_i$, and the Jacobians of the power flow equations collected in $C_i$ are of dimension $(n_i/2) \times n_i$.
Recall that $n_i = 4 N_i$, where $N_i$ is the number of buses in region $i$, cf. Section~\ref{sec:ACOPF_separable}.
Additionally, the vector of binaries indicating the active bounds $\mathbb{A}_i$, which are of dimension $\frac{3 n_i}{4}$, has to be communicated (bounds on power injections and voltages).
Hence, the worst case forward communication need for \aladin comprises
$
\sum_{i\in \mathcal{R}}  {n_{i} + \frac{n_{i} (n_{i}+1)}{2}+ \frac{n_{i}^2}{2}} = \sum_{i\in \mathcal{R}} \frac{n_i (2 n_i + 3)}{2}
$
floats and ${\frac{3 n_i}{4}}$ binaries. }

\comm{The block-\bfgs update described in Section~\ref{sec:ALADIN} reduces the total communication need as follows.
Instead of having to communicate the Hessians $H_i$ which lead to the quadratic term $\frac{n_{i} (n_{i}+1)}{2}$, \bfgs requires to communicate only the $n_i$-dimensional gradients of the Lagrangian.
Hence, the worst-case forward communication need reduces to
$
\sum_{i\in \mathcal{R}}  {n_{i} + n_i + \frac{n_{i}^2}{2}} = \sum_{i\in \mathcal{R}} \frac{n_i (n_i +4)}{2}
$
floats and $\frac{3 n_i}{4}$ binaries.}

\comm{After solving  \qp \eqref{eq::conqp}, primal and dual steps for the consensus constraint are broadcasted to the subproblems. 
The number of consensus constraints should typically be smaller than the decision variables since otherwise the original problem might be infeasible. 
Hence, the number of Lagrange multipliers is upper-bounded by $\sum_{i \in \mathcal{R}}n_i$, and we obtain an upper bound for the backward communication effort of
$
\sum_{i \in \mathcal{R}}n_{i} + n_i
$
floats.}

For \admm, only the minimizers of the local problems have to be communicated in both directions. As a result, we obtain equal worst case forward and backward communication need of
$
\sum_{i\in \mathcal{R}} n_{i}
$
floats.

\comm{Table~\ref{tab::comparisonAll} summarizes the results of this section, comparing convergence properties, convergence rates, and communication effort in terms of floats for \admm and both variants of \aladin. 
Table~\ref{tab::comparisonAll} introduces the short-hand notations $\nFW$ ($\nBW$) for the worst case forward (backward) communication effort in terms of floats.}

\begin{remark}[Floats vs. Binaries]
Observe that communicating a binary value is much cheaper than communicating floats   {(1~bit~vs.~32~or~64 bits)}. Hence, counting the floats is usually sufficient to approximately determine communication effort. 
\end{remark}

\subsubsection{Communication Effort in Practice}\label{sec:CommEffPrac}
\comm{
In practice, the Hessian and Jacobian approximations often contain many structural zeros.
If the central coordinator knows the sparsity pattern, these zeros do not have to be communicated. Table~\ref{tab::Comm&timing} compares the upper bounds derived above to the worst case per step communication effort occuring in our simulations counting the maximum number of non-zero floats during all iterations. One can observe that the  communication effort is approximately a factor of four smaller in practice compared with their upper bounds. More precisely,  
  in Table~\ref{tab::Comm&timing} we observe $ \nFWeff <  \nFW$, where  $ \nFWeff$ is the forward communication effort in our simulations and $\nFW$ is the upper bound. Furthermore, the communication overhead for \aladin is larger compared with \admm---both per step and in the total communication effort.
The use of \bfgs reduces the communication effort by at least a factor of two. Generally one can say that the reduction factor gained by \bfgs grows with increasing problems caused by the quadratic growth of the number of variables in the Hessian with problem size. }

\subsubsection{Worst Case Computation Time}
\comm{
	Next, we assess worst case computation times for \aladin and \admm.
	Note that structurally the local \nlps are the same for \aladin and \admm.
	Let $\tNLP$ denote the worst-case time to solve any of the local~\nlps in any iteration using \aladin.
	For the coordination step, \aladin requires additional time to solve the \qp (denoted by $\tQP$), while we assume that the averaging time for \admm is negligible.
	In order to enable a fair comparison we introduce the worst case computation time as follows
	\begin{equation}
	\label{eq:WorstCaseComputationTime}
	\tWC {=} 
	\begin{cases}
			\text{\#Iterations} \cdot  \tNLP, & \text{for \admm}, \\
			\text{\#Iterations} \cdot  (\tNLP + \tQP), & \text{for \aladin}.
	\end{cases}
	\end{equation}
	 We bound the time needed by \admm to solve  the local \nlps by $\tNLP$ obtained via our numerical \aladin experiments. This way we intend to focus on the algorithmic differences between \aladin and \admm and not on the details of specific implementations.}
	 
\comm{	Hence, \aladin needs more time per iteration, but---given the faster convergence of \aladin from Table \ref{tab::comparison}---\aladin still outperforms \admm in terms of the worst case computation time.
	In fact, the total worst case computation time for \aladin is at least a factor of two smaller compared with \admm.
	Table~\ref{tab::Comm&timing} shows the worst case computation times for the test cases.}

\section{Conclusion \& Outlook}
This paper investigated the potential of applying the  Augmented Lagrangian Alternating Direction Inexact Newton (\aladin) method to distributed \ac-\opf problems. 
The presented numerical results for grids of different sizes illustrate the potential of \aladin for \ac-\opf. In comparison with \admm, \aladin is able to reduce the number of iterations by at least one order of magnitude. This comes at the cost of an increased per-step communication effort which can be reduced by using inexact Hessians, for example via \bfgs updates. Doing so, we increase the number of iterations slightly  but \aladin remains faster and more accurate than \admm.

While the present paper focused primarily on comparing \aladin with \admm, a detailed comparison with other distributed schemes will be of interest. Moreover, future work will consider multi-stage \opf problems including storages and generator ramp constraints. From an algorithmic and communication point of view, it seems promising to reduce the communication effort even more, e.g. by formulating the coordination \qp in the coupling variables only. \comm{The development of  improved (distributed) line search strategies and performing tests on larger grids including sensitivity analysis to grid topology and load patterns is subject of ongoing and future work.}

\newpage
\appendix

\label{sec:LocConvProof}

\begin{proof}
	From Assumption~\ref{ass:TechnicalConditions}, item~\ref{ass::inesol} we have 
	\begin{eqnarray}\notag
	\label{eq::ineq1}
	\|\overline{x}^k - x^\star\| &\leq& \|\overline{x}^k - x^k\| + \|x^k - x^\star\|\\
	&\leq& \zeta_1\|z^k - x^k \| + \|x^k - x^\star\|\\\notag
	&\leq& \zeta_1\|z^k - x^\star\| + (\zeta_1+1)\|x^k - x^\star\|.
	\end{eqnarray}
	From~\cite[Lem. 3]{Houska2016}, we know that there exist constants $\zeta_2,\zeta_3>0$ such that the solutions of~\eqref{eq::denlp} satisfy
	\begin{equation}
	\label{eq::ineq2}
	\|x^k - x^\star\| \leq \zeta_2\|z^k - x^\star\| + \zeta_3\|\lambda^k -\lambda^\star\|.
	\end{equation}
	Combining~\eqref{eq::ineq1} and~\eqref{eq::ineq2} yields
	\begin{eqnarray}
	\label{eq::ineq3}
	\|\overline{x}^k - x^\star\| \leq \omega_1\|z^k - x^\star\| + \omega_2\|\lambda^k -\lambda^\star\|.
	\end{eqnarray}
	with $\omega_1 = \zeta_1 + (\zeta_1+1)\zeta_2$ and $\omega_2 = (\zeta_1+1)\zeta_3$. If we use  exact Hessians and Jacobians and $\mu^k$ satisfies~\eqref{eq::mu}, then there exists a constant $0<\omega_3<\infty$ such that 
	\[
	\|z^{k+1} - x^\star \|\leq \frac{\omega_3}{2}\|\overline{x}^k - x^\star\|^2\;,\;\|\lambda^{k+1} - \lambda^\star \|\leq \frac{\omega_3}{2}\|\overline{x}^k - x^\star\|^2,
	\]
	see~\cite{Houska2016}. Here, we use that $(x^*,w^*)$ is a regular \kkt point which yields in combination with~\eqref{eq::ineq3}
	\begin{multline}
	\label{eq:AppInequality}
	 \|z^{k+1} - x^\star \| +  \|\lambda^{k+1}- \lambda^\star \|\\
	\leq \omega_3 \left(\omega_1\|z^k - x^\star\| + \omega_2\|\lambda^k -\lambda^\star\| \right)^2.
	\end{multline}
	The above inequality allows concluding a quadratic convergence rate as $\omega_1$ and $\omega_2$ are strictly positive and finite.	
\end{proof}

\section{Partitioning Data and \aladin Settings}
\begin{table}
	\caption{Parameterization of \aladin for shown \ieee test cases. The unit of $\gamma$ is $\$/\text{hr}/(p.u.)^2$.}
	\centering
	\renewcommand{\arraystretch}{1.7}
	\begin{tabular}{ccccccccccccccccccc}
		\toprule
	Test Case &  $\underline \rho$ & $\overline \rho$ & $r_\rho$& $\underline \mu$  & $\overline \mu$ & $r_\mu$& $\gamma$ & \\ 
		\midrule
		5   & 10\textsuperscript{2} & 10\textsuperscript{6} & 1.5& 10\textsuperscript{3} & 2$\cdot$10\textsuperscript{6} & 2& 0  \\	
		30  & 10\textsuperscript{2} & 10\textsuperscript{6} & 1.5& 10\textsuperscript{3} & 2$\cdot$10\textsuperscript{6} & 2&10\\	
		57  & 10\textsuperscript{2} & 10\textsuperscript{6} & 1.5& 10\textsuperscript{3} & 2$\cdot$10\textsuperscript{6} & 2&10\\	
		118 & 10\textsuperscript{2} & 10\textsuperscript{6} & 1.1& 10\textsuperscript{3} & 2$\cdot$10\textsuperscript{6} & 2&10\\	
		300 & 10\textsuperscript{2} & 10\textsuperscript{3} & 0.8& 10\textsuperscript{3} & 2$\cdot$10\textsuperscript{6} & 2&0  \\
		\bottomrule		 	
	\end{tabular}
	\label{tab::prameters}
\end{table}

\begin{table}
	\caption{Grid partitioning (excluding auxiliary buses).}
	\centering
	\renewcommand{\arraystretch}{1.7}
	\begin{tabular}{lll}
		\toprule
		Test Case	& $|\mathcal{A}|$ & Regions $\mathcal{N}_i$  \\ \midrule
		5  & 4 & \{1, 5\}, \{2, 3\}, \{4\}\\ 
		30 & 8 & \{1--8, 28\}, \{9--11, 17, 21, 22\}\\
		&& \{24--27, 29, 30\}, \{12--16, 18-20, 23\}\\
		57 & 24 & \{24--26, 30--33\}, \{10, 12, 16, 17, 51\}, \\ && \{8, 9, 11, 41--43, 55--57\}  \{13, 14, 46--50\}, \\&&
		 \{34--37, 39, 40\},  \{7, 27--29, 52--54\},\\&& \{19--23, 38, 44\}, \{1--6, 15, 18, 45\}\\
		118& 13 & \{1--32, 113--115, 117\}, \{33--67\}, \\&& \{68--81, 116, 118\}, \{82, 112\}\\
		300& 61 & \{1--100\}, \{101--200\}, \{201--300\}\\	
		\bottomrule
	\end{tabular}
	\label{tab::partitioning}
\end{table}

%
%

\ifCLASSOPTIONcaptionsoff
  \newpage
\fi



%
%
%

\printbibliography

%







\end{document}